%% file: main.tex
\pdfoutput=1
\documentclass[12pt,a4paper]{article}

\usepackage{ifthen} 
\newboolean{pdflatex}
\setboolean{pdflatex}{true} 

\newboolean{articletitles}
\setboolean{articletitles}{true} 

\newboolean{uprightparticles}
\setboolean{uprightparticles}{false} 

\def\paperauthors{LHCb collaboration} 
\def\paperasciititle{Modification of chi_c1(3872) and psi(2S) production in pPb collisions at sqrt(s_NN) = 8.16 TeV} 
\def\papertitle{Modification of $\theX$ and $\psitwos$ production in $p$Pb collisions at ${\ensuremath{\protect\sqrt{s_{\scriptscriptstyle\text{NN}}}}\xspace} = 8.16\tev$}
\def\paperkeywords{{LHCb}} 
\def\papercopyright{\the\year\ CERN for the benefit of the LHCb collaboration} 
\def\paperlicence{CC BY 4.0 licence}
\def\paperlicenceurl{https://creativecommons.org/licenses/by/4.0/}

\input{preamble}
\usepackage{longtable} 
\usepackage{placeins}
\begin{document}

\renewcommand{\thefootnote}{\fnsymbol{footnote}}
\setcounter{footnote}{1}

\input{title-LHCb-PAPER.tex}


\renewcommand{\thefootnote}{\arabic{footnote}}
\setcounter{footnote}{0}

\cleardoublepage


\pagestyle{plain} 
\setcounter{page}{1}
\pagenumbering{arabic}


\input{body}

\input{acknowledgements}




\addcontentsline{toc}{section}{References}
\bibliographystyle{LHCb}
\bibliography{main,standard,LHCb-PAPER,LHCb-CONF,LHCb-DP,LHCb-TDR}

\newpage
\input{Authorship_LHCb-PAPER-2023-026}

\end{document}

%% file: preamble.tex
\usepackage[top=1in, bottom=1.25in, left=1in, right=1in]{geometry}

\usepackage{microtype}
\usepackage{lineno}  
\usepackage{xspace} 
\usepackage{caption} 

\usepackage{graphicx}  
\usepackage{color}
\usepackage{colortbl}
\graphicspath{{./figs/}} 

\usepackage{amsmath} 
\usepackage{amssymb}
\usepackage{amsfonts}
\usepackage{upgreek} 

\newcommand*\patchAmsMathEnvironmentForLineno[1]{%
\expandafter\let\csname old#1\expandafter\endcsname\csname #1\endcsname
\expandafter\let\csname oldend#1\expandafter\endcsname\csname
end#1\endcsname
 \renewenvironment{#1}%
   {\linenomath\csname old#1\endcsname}%
   {\csname oldend#1\endcsname\endlinenomath}%
}
\newcommand*\patchBothAmsMathEnvironmentsForLineno[1]{%
  \patchAmsMathEnvironmentForLineno{#1}%
  \patchAmsMathEnvironmentForLineno{#1*}%
}
\AtBeginDocument{%
\patchBothAmsMathEnvironmentsForLineno{equation}%
\patchBothAmsMathEnvironmentsForLineno{align}%
\patchBothAmsMathEnvironmentsForLineno{flalign}%
\patchBothAmsMathEnvironmentsForLineno{alignat}%
\patchBothAmsMathEnvironmentsForLineno{gather}%
\patchBothAmsMathEnvironmentsForLineno{multline}%
\patchBothAmsMathEnvironmentsForLineno{eqnarray}%
}

\usepackage{hyperxmp}

\usepackage[pdftex,
            pdfauthor={\paperauthors},
            pdftitle={\paperasciititle},
            pdfkeywords={\paperkeywords},
            pdfcopyright={Copyright (C) \papercopyright},
            pdflicenseurl={\paperlicenceurl}]{hyperref}

\usepackage[colorinlistoftodos,textsize=scriptsize]{todonotes}

\usepackage[bottom,flushmargin,hang,multiple]{footmisc}

\usepackage[all]{hypcap} 

\input{lhcb-symbols-def} 

\usepackage{cite} 
\usepackage{mciteplus}

%% file: lhcb-symbols-def.tex
\usepackage{xspace} 
\usepackage{upgreek}

\def\lhcb   {\mbox{LHCb}\xspace}

\def\MagUp {\mbox{\em Mag\kern -0.05em Up}\xspace}

\ifthenelse{\boolean{uprightparticles}}%
{

 \def\Pmu         {\ensuremath{\upmu}\xspace}

 \def\Ppi         {\ensuremath{\uppi}\xspace}

 \def\Pchi        {\ensuremath{\upchi}\xspace}                 
 \def\Ppsi        {\ensuremath{\uppsi}\xspace}

 \def\PDelta      {\ensuremath{\Delta}\xspace}                 
 \def\PXi         {\ensuremath{\Xi}\xspace}                 
 \def\PLambda     {\ensuremath{\Lambda}\xspace}                 
 \def\PSigma      {\ensuremath{\Sigma}\xspace}                 
 \def\POmega      {\ensuremath{\Omega}\xspace}                 
 \def\PUpsilon    {\ensuremath{\Upsilon}\xspace}
 \let\oldPi\Pi
 \def\PPi         {\ensuremath{\oldPi}\xspace}

 \def\PB      {\ensuremath{\mathrm{B}}\xspace}                 
                  
 \def\PD      {\ensuremath{\mathrm{D}}\xspace}

 \def\PJ      {\ensuremath{\mathrm{J}}\xspace}                 
 \def\PK      {\ensuremath{\mathrm{K}}\xspace}

 \def\Pb      {\ensuremath{\mathrm{b}}\xspace}

 \def\Pi      {\ensuremath{\mathrm{i}}\xspace}

 \def\Ps      {\ensuremath{\mathrm{s}}\xspace}

 \def\thebaroffset{0.0em}
}
{

 \def\Pmu         {\ensuremath{\mu}\xspace}

 \def\Ppi         {\ensuremath{\pi}\xspace}

 \def\Pchi        {\ensuremath{\chi}\xspace}                 
 \def\Ppsi        {\ensuremath{\psi}\xspace}                 
                  
 \mathchardef\PDelta="7101
 \mathchardef\PXi="7104
 \mathchardef\PLambda="7103
 \mathchardef\PSigma="7106
 \mathchardef\POmega="710A
 \mathchardef\PUpsilon="7107
 \mathchardef\PPi="7105
                  
 \def\PB      {\ensuremath{B}\xspace}                 
                  
 \def\PD      {\ensuremath{D}\xspace}

 \def\PJ      {\ensuremath{J}\xspace}                 
 \def\PK      {\ensuremath{K}\xspace}

 \def\Pb      {\ensuremath{b}\xspace}

 \def\Pi      {\ensuremath{i}\xspace}

 \def\Ps      {\ensuremath{s}\xspace}

 \def\thebaroffset{0.18em}
}
\newcommand{\offsetoverline}[2][\thebaroffset]{\kern #1\overline{\kern -#1 #2}}%

\makeatletter
\ifcase \@ptsize \relax
  \newcommand{\miniscule}{\@setfontsize\miniscule{4}{5}}
\or
  \newcommand{\miniscule}{\@setfontsize\miniscule{5}{6}}
\or
  \newcommand{\miniscule}{\@setfontsize\miniscule{5}{6}}
\fi
\makeatother

\DeclareRobustCommand{\optbar}[1]{\shortstack{{\miniscule (\rule[.5ex]{1.25em}{.18mm})}
  \\ [-.7ex] $#1$}}

\def\mup        {{\ensuremath{\Pmu^+}}\xspace}

\def\squark    {{\ensuremath{\Ps}}\xspace}

\def\bquark    {{\ensuremath{\Pb}}\xspace}

\def\pion   {{\ensuremath{\Ppi}}\xspace}

\def\pip    {{\ensuremath{\pion^+}}\xspace}
\def\pim    {{\ensuremath{\pion^-}}\xspace}

\def\KorKbar {\kern \thebaroffset\optbar{\kern -\thebaroffset \PK}{}\xspace}

\def\D       {{\ensuremath{\PD}}\xspace}

\def\DorDbar {\kern \thebaroffset\optbar{\kern -\thebaroffset \PD}\xspace}

\def\Dp      {{\ensuremath{\D^+}}\xspace}
\def\Dm      {{\ensuremath{\D^-}}\xspace}

\def\DpDm    {\ensuremath{\Dp {\kern -0.16em \Dm}}\xspace}

\def\B       {{\ensuremath{\PB}}\xspace}

\def\BorBbar {\kern \thebaroffset\optbar{\kern -\thebaroffset \PB}\xspace}

\def\Bd      {{\ensuremath{\B^0}}\xspace}

\def\BdorBdbar {\kern \thebaroffset\optbar{\kern -\thebaroffset \Bd}\xspace}

\def\Bs      {{\ensuremath{\B^0_\squark}}\xspace}

\def\BsorBsbar {\kern \thebaroffset\optbar{\kern -\thebaroffset \Bs}\xspace}

\def\jpsi     {{\ensuremath{{\PJ\mskip -3mu/\mskip -2mu\Ppsi}}}\xspace}
\def\psitwos  {{\ensuremath{\Ppsi{(2S)}}}\xspace}

\def\Y#1S{\ensuremath{\PUpsilon{(#1S)}}\xspace}

\def\theX     {{\ensuremath{\Pchi_{c1}(3872)}}\xspace}

\def\LorLbar     {\kern \thebaroffset\optbar{\kern -\thebaroffset \PLambda}\xspace}

\def\BF         {{\ensuremath{\mathcal{B}}}\xspace}
\def\BR         {\BF}

\def\AT#1     {\ensuremath{A_{\mathrm{T}}^{#1}}\xspace}           

\def\C#1      {\ensuremath{\mathcal{C}_{#1}}\xspace}                       
\def\Cp#1     {\ensuremath{\mathcal{C}_{#1}^{'}}\xspace}                    
\def\Ceff#1   {\ensuremath{\mathcal{C}_{#1}^{\mathrm{(eff)}}}\xspace}        
\def\Cpeff#1  {\ensuremath{\mathcal{C}_{#1}^{'\mathrm{(eff)}}}\xspace}       
\def\Ope#1    {\ensuremath{\mathcal{O}_{#1}}\xspace}                       
\def\Opep#1   {\ensuremath{\mathcal{O}_{#1}^{'}}\xspace}                    

\newcommand{\aunit}[1]{\ensuremath{\text{\,#1}}}       

\newcommand{\tev}{\aunit{Te\kern -0.1em V}\xspace}
\newcommand{\gev}{\aunit{Ge\kern -0.1em V}\xspace}
\newcommand{\mev}{\aunit{Me\kern -0.1em V}\xspace}
\newcommand{\kev}{\aunit{ke\kern -0.1em V}\xspace}
\newcommand{\ev}{\aunit{e\kern -0.1em V}\xspace}
 
\newcommand{\mevc}{\ensuremath{\aunit{Me\kern -0.1em V\!/}c}\xspace}
\newcommand{\gevc}{\ensuremath{\aunit{Ge\kern -0.1em V\!/}c}\xspace}
\newcommand{\mevcc}{\ensuremath{\aunit{Me\kern -0.1em V\!/}c^2}\xspace}
\newcommand{\gevcc}{\ensuremath{\aunit{Ge\kern -0.1em V\!/}c^2}\xspace}

\def\gsim{{~\raise.15em\hbox{$>$}\kern-.85em
          \lower.35em\hbox{$\sim$}~}\xspace}
\def\lsim{{~\raise.15em\hbox{$<$}\kern-.85em
          \lower.35em\hbox{$\sim$}~}\xspace}

\def\sPlot{\mbox{\em sPlot}\xspace}

\def\sqs   {\ensuremath{\protect\sqrt{s}}\xspace}
\def\sqsnn {\ensuremath{\protect\sqrt{s_{\scriptscriptstyle\text{NN}}}}\xspace}
\def\pt         {\ensuremath{p_{\mathrm{T}}}\xspace}

\def\evtgen     {\mbox{\textsc{EvtGen}}\xspace}

\def\geant      {\mbox{\textsc{Geant4}}\xspace}

\def\pythia     {\mbox{\textsc{Pythia}}\xspace}

\def\tell1  {TELL1\xspace}
\def\ukl1   {UKL1\xspace}

\newcommand{\lhcborcid}[1]{\href{https://orcid.org/#1}{\hspace*{0.1em}\raisebox{-0.45ex}{\includegraphics[width=1em]{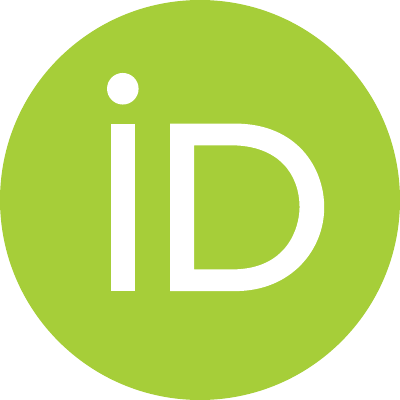}}}}


%% file: title-LHCb-PAPER.tex

\begin{titlepage}
\pagenumbering{roman}

\vspace*{-1.5cm}
\centerline{\large EUROPEAN ORGANIZATION FOR NUCLEAR RESEARCH (CERN)}
\vspace*{1.5cm}
\noindent
\begin{tabular*}{\linewidth}{lc@{\extracolsep{\fill}}r@{\extracolsep{0pt}}}
\ifthenelse{\boolean{pdflatex}}
{\vspace*{-1.5cm}\mbox{\!\!\!\includegraphics[width=.14\textwidth]{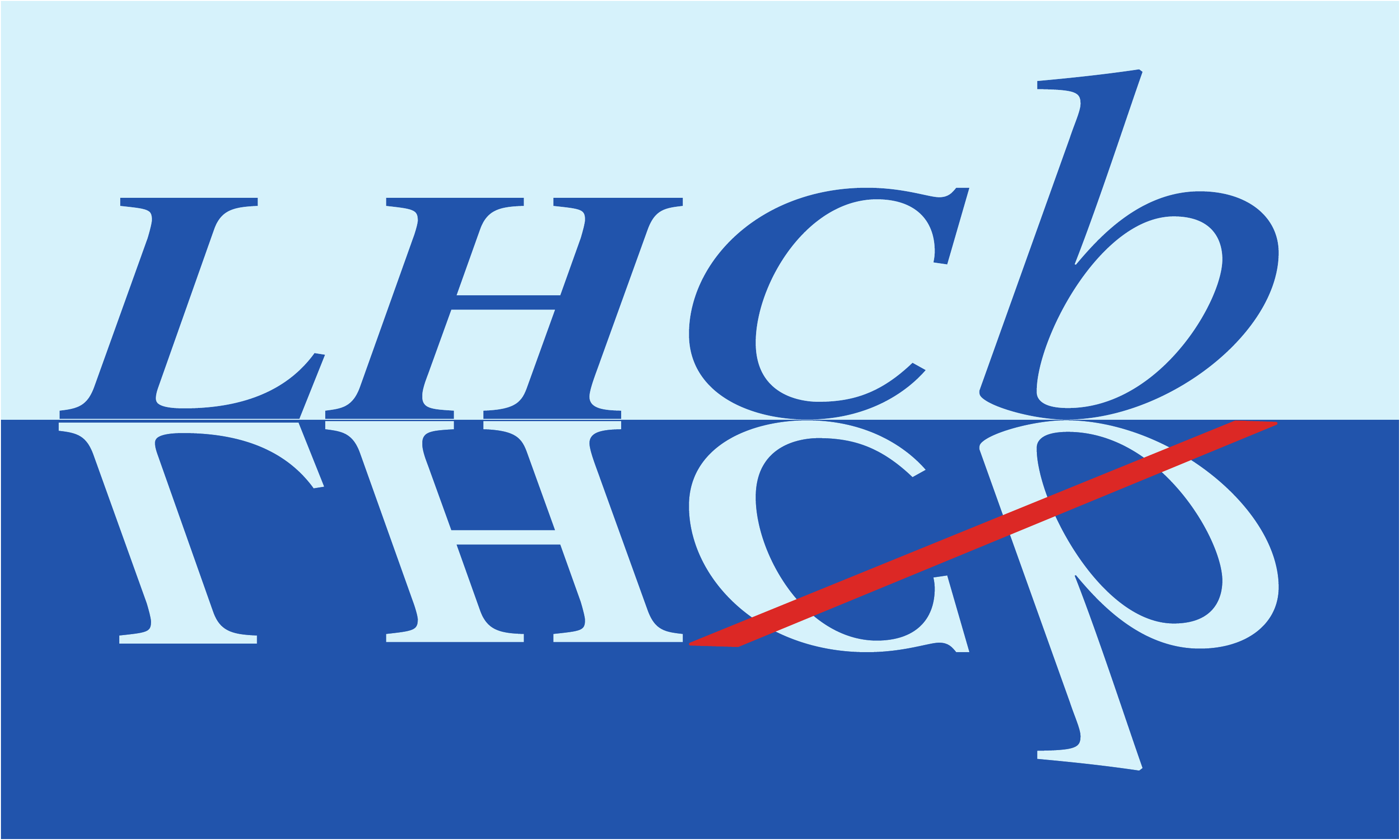}} & &}%
{\vspace*{-1.2cm}\mbox{\!\!\!\includegraphics[width=.12\textwidth]{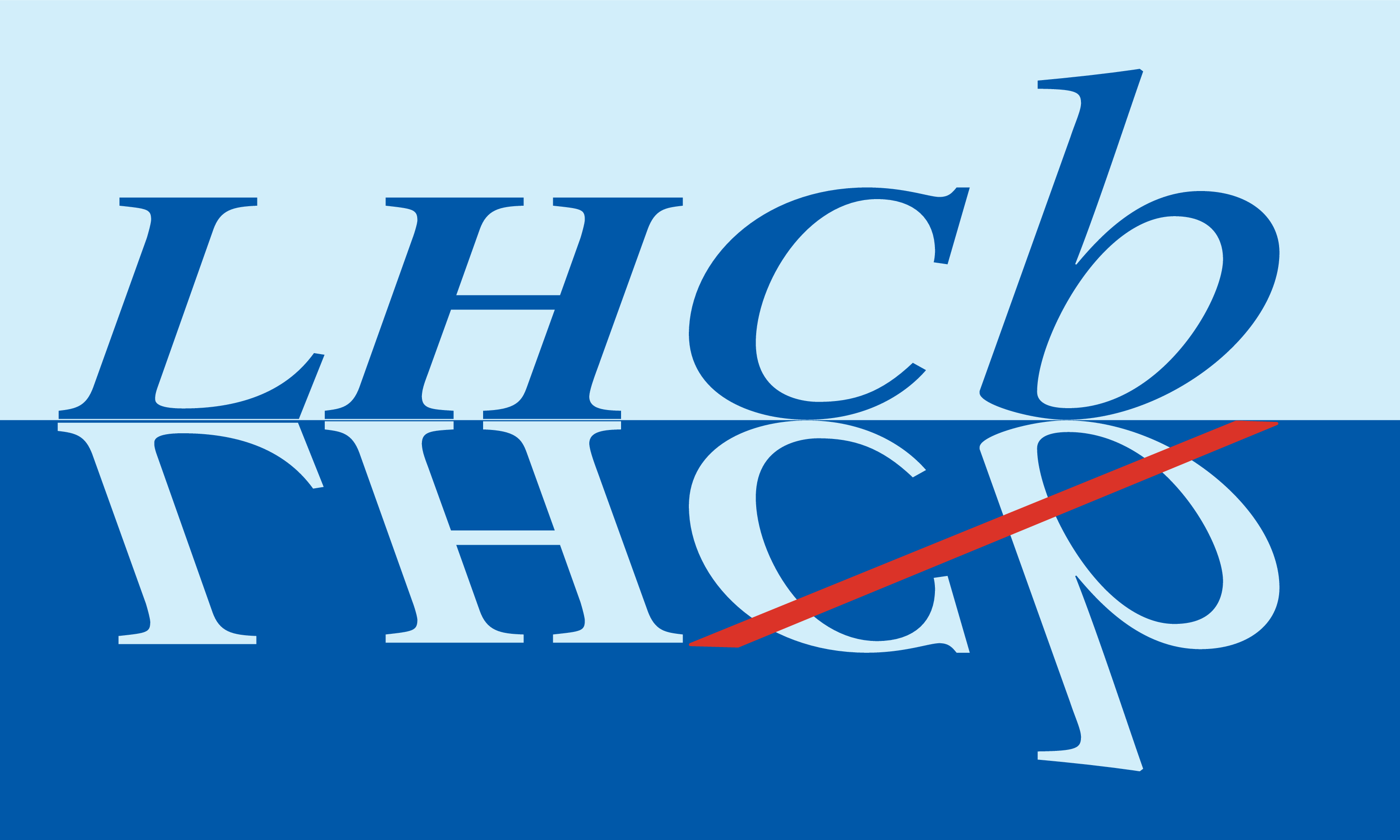}} & &}%
\\
 & & CERN-EP-2024-033 \\  
 & & LHCb-PAPER-2023-026 \\  
 & & June 19, 2024 \\ 
 & & \\
\end{tabular*}

\vspace*{4.0cm}

{\normalfont\bfseries\boldmath\huge
\begin{center}
  \papertitle 
\end{center}
}

\vspace*{2.0cm}

\begin{center}
\paperauthors\footnote{Authors are listed at the end of this paper.}
\end{center}

\vspace{\fill}

\begin{abstract}
  \noindent
The LHCb collaboration measures production of the exotic hadron $\theX$ in proton-nucleus collisions for the first time.  Comparison with the charmonium state $\psitwos$ suggests that the exotic $\theX$ experiences different dynamics in the nuclear medium than conventional hadrons, and comparison with data from proton-proton collisions indicates that the presence of the nucleus may modify $\theX$ production rates.  This is the first measurement of the nuclear modification factor of an exotic hadron.

\end{abstract}

\vspace*{2.0cm}

\begin{center}
  Published in Phys. Rev. Lett. 132 (2024) 242301
\end{center}

\vspace{\fill}

{\footnotesize 
\centerline{\copyright~\papercopyright. \href{\paperlicenceurl}{\paperlicence}.}}
\vspace*{2mm}

\end{titlepage}


\newpage
\setcounter{page}{2}
\mbox{~}
%
%
%
%

%% file: body.tex
The study of exotic hadrons with more than three valence quarks is a highly active area of quantum chromodynamics.  Dozens of exotic states have been discovered in the last 20 years, and various models such as compact tetraquarks, hadronic molecules, hadrocharmonia, and other structures have been proposed in attempts to explain their various properties (see reviews in Refs.~\cite{Olsen:2017bmm,Guo:2017jvc,Chen:2016qju,Lebed:2016hpi}).  However, to date, there is no general consensus on the nature of the first discovered and most well-studied exotic hadron, the $\theX$ state.

Most existing measurements of the properties of exotic hadrons containing charm quarks utilize their production in the decays of hadrons containing $\bquark$ quarks.  These decays provide well-defined initial conditions, and many sources of background can be efficiently rejected using the relatively long lifetime of $b$ hadrons. The LHCb experiment has used these data samples to obtain precise measurements of $\theX$ properties such as its quantum numbers, mass, and width~\cite{LHCb-PAPER-2013-001,LHCb-PAPER-2015-015,LHCb-PAPER-2020-008,LHCb-PAPER-2020-009}, and to explore new $\theX$ production channels and decays~\cite{LHCb-PAPER-2014-008,LHCb-PAPER-2019-023,LHCb-PAPER-2021-045, LHCb-PAPER-2022-049}.  However, exotic hadrons can also be produced promptly at the interaction point of hadronic collisions, where they can interact with other particles produced in the event.  In collisions using beams of nuclei, exotic hadrons can also interact with the nuclear remnant and may be subject to the effects of quark-gluon plasma.  The response of the exotic hadrons to these effects provides new ways to constrain their properties, which are not accessible when studying $\bquark$-hadron decays.

Previous measurements by the LHCb collaboration in $pp$ collisions showed a significant decrease in the ratio of prompt $\theX$ to $\psitwos$ cross-sections, $\sigma^{\theX}/\sigma^{\psitwos}$, with increasing charged-particle multiplicity~\cite{LHCb-PAPER-2020-023}.  These data were interpreted in terms of breakup of the $\theX$ hadrons due to interactions with comoving particles produced in the event, for both compact and molecular models of $\theX$ structure~\cite{Esposito:2020ywk,Braaten:2020iqw}.  The CMS collaboration has measured the $\sigma^{\theX}/\sigma^{\psitwos}$ ratio in PbPb collisions, and found that the ratio is enhanced relative to $pp$ collisions, although that measurement has large uncertainties~\cite{CMS:2021znk}. Statistical hadronization models predict that $\theX$ production is significantly enhanced in PbPb collisions at the LHC~\cite{x3872_enhancement,andronic_shm}.  Calculations based on quark coalescence, which can occur when quark wavefunctions overlap in position and velocity space, show that production rates of $\theX$ hadrons in nucleus-nucleus (AA) collisions are sensitive to its structure.  In these models, production of compact tetraquarks is expected to be greatly enhanced over hadronic molecules~\cite{exhic_1,exhic_2}, although a recent transport calculation reaches the opposite conclusion~\cite{Wu:2020zbx}.  Late-stage interactions in the hadron gas phase of a heavy-ion collision can also affect the observed yields~\cite{X3872_hadrongas}.  The suppressing effects of breakup and the enhancing effects of coalescence are expected to dominate in different multiplicity regimes~\cite{Guo:2023dwf}, and it is currently unknown where the crossover may occur.

Collisions of protons with Pb nuclei provide an intermediate stage between the relatively small $pp$ collision system and the large PbPb system, and can thereby shed light on the interplay of various enhancement and suppression mechanisms.  Calculations of tetraquark production in $p$Pb collisions have predicted that the $\theX$ cross-section could be enhanced relative to $pp$ collisions, due to a higher rate of double-parton scattering~\cite{Carvalho:2017mdi}.  An increase of double-parton scattering in $p$Pb collisions relative to $pp$ collisions has since been measured by the LHCb collaboration~\cite{LHCb-PAPER-2020-010}.  An enhancement of proton production relative to pions and kaons has been observed in $d$Au and $p$Pb collisions~\cite{STAR:2003oii, PHENIX:2013kod, ALICE:2016dei}, which can be explained by coalescence of three quarks into baryons versus two quarks into mesons~\cite{Hwa:2004zd,Hwa:2004yi,Shao:2017eok}.  Similarly, an enhancement of charmed baryons relative to charmed mesons has been observed in $pp$ and $p$Pb collisions, relative to expectations from $e^{+}e^{-}$ collisions~\cite{ALICE:2021dhb,ALICE:2020wfu}, which may be explained by quark coalescence.  These coalescence effects could be even more pronounced for four-quark states, which have not previously been measured in $p$A collisions.  Therefore, in addition to providing novel information on the $\theX$ structure, measurements in $p$Pb collisions can provide new tests of models of particle transport and hadronization in nuclear collisions, in a new range of number of constituent quarks.

This Letter describes the first measurements of the prompt production of the exotic state $\theX$ in $p$Pb collisions, including the ratio of $\theX$ to $\psitwos$ cross-sections and the $\theX$ nuclear modification factor $R_{p \rm A}^{\theX}$.  The $\theX$ and $\psitwos$ hadrons are reconstructed through their decays to $\jpsi \pip \pim$, where the $\jpsi$ particle subsequently decays to a pair of oppositely charged muons.  These measurements use $pp$ and $p$Pb collision data recorded by the LHCb experiment.  
The $pp$ data were collected in 2012 at a center-of-mass energy $\sqs = 8\tev$, corresponding to an integrated luminosity of about 2 fb$^{-1}$.  The $p$Pb data were collected in 2016 in two configurations. In the forward configuration, denoted $p$Pb, the proton beam is directed into the LHCb spectrometer and measurements cover the rapidity interval $1.5<y<4$, where $y$ is measured in the center-of-mass frame of the proton-nucleus system.  In the backward configuration, denoted Pb$p$, the Pb beam travels into the spectrometer and the resulting rapidity coverage is $-5<y<-2.5$.  The $p$Pb and Pb$p$ data sets considered here were recorded at a center-of-mass energy per nucleon pair $\sqsnn = 8.16\tev$, and correspond to integrated luminosities of about 12.5 and 19.3 nb$^{-1}$, respectively.

The LHCb detector is a single-arm forward spectrometer, described in detail in Refs.~\cite{LHCb-DP-2008-001,LHCb-DP-2014-002}.  Events considered in this analysis are selected with a series of triggers which retain events containing the decay $\jpsi\rightarrow\mup\mu^{-}$.  The offline selection requires muon candidates to have total momentum $p > 3\gevc$ and transverse momentum $\pt> 650\mevc$, and to penetrate hadron absorbers in the muon system.  Candidate $\jpsi$ mesons are formed from pairs of oppositely charged muon candidates that have an invariant mass within three standard deviations ($\sim39\mevcc$) of the mean of the $\jpsi$ peak.  Charged pion candidates are required to have $p>3\gevc$ and $\pt > 500\mevc$, and are identified by the response of the ring-imaging Cherenkov detectors.  Combinations of $\mup\mu^{-}\pip\pim$ candidates that form a good quality common vertex are retained, and the tracks are refit with kinematic constraints that require all four tracks to originate from a common vertex and constrain the $\mup\mu^{-}$ invariant mass to the known $\jpsi$ mass~\cite{PDG2022}.  The difference between the $\jpsi\pim\pip$ mass and the sum of the $\jpsi$ and $\pip\pim$ masses is required to be less than $300\mevcc$, which reduces combinatorial backgrounds while retaining signal.  The resulting $\jpsi\pip\pim$ candidates are required to have $\pt>5\gevc$.

\begin{figure}[!hb]
  \centering
 \includegraphics[width=0.8\linewidth]{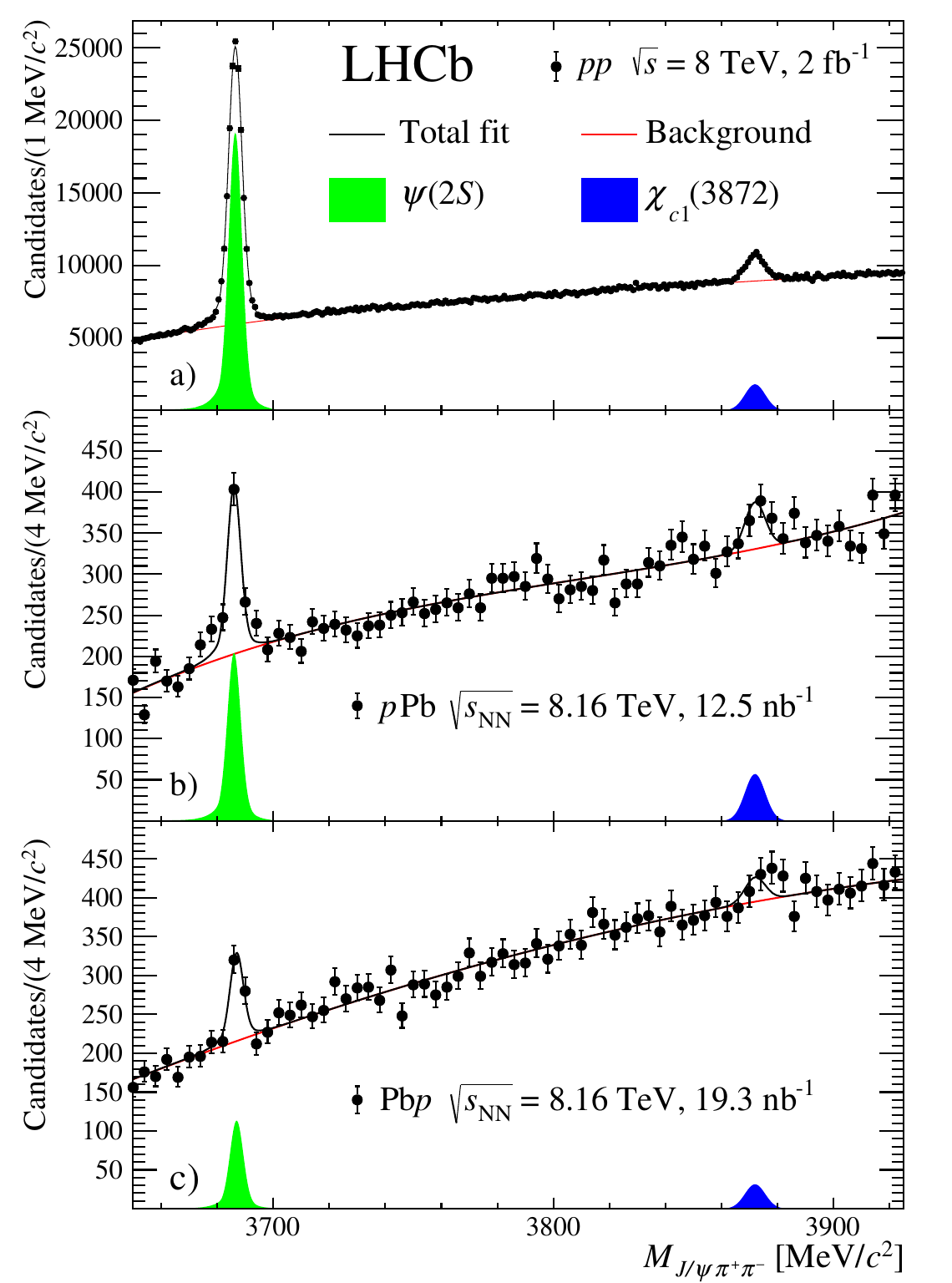}
  \caption{Invariant mass spectra of $\jpsi \pip \pim$ candidates measured in a) $pp$, b) $p$Pb, and c) Pb$p$ collisions, with fit projections overlaid.}
  \label{fig:Fig1}
\end{figure}

The $\theX$ and $\psitwos$ signals of interest are produced promptly at the collision vertex, where they are subject to interactions with other particles in the event.  The pseudo decay-time $t_z$ is used to select promptly produced signal candidates and reject those produced in decays of $b$ hadrons.  This variable is defined as

\begin{equation}
t_z  \equiv \frac{ (z_{\rm decay} - z_{\rm PV}) \times M } { p_{z} },
\end{equation}

\noindent where $z_{\rm decay} - z_{\rm PV}$ is the difference between the positions of the reconstructed vertex of the $\jpsi \pip \pim$ candidate and the associated collision vertex along the beam axis, $M$ is the mass of the reconstructed signal candidate, and $p_z$ is the candidate's momentum along the beam axis.  A requirement of $t_z<0.1$ ps is applied.  The data and simulations show that this retains more than $99\%$ of the prompt signals, while rejecting $\sim80\%$ of the signals produced in decays of $\bquark$ hadrons.  Previous measurements have shown that the fraction of $\theX$ and $\psitwos$ that are produced promptly in $pp$ collisions at 8 TeV are about $80\%$ and $75\%$, respectively,~\cite{LHCb-PAPER-2020-023}, and that $b$ hadron production is not significantly modified in $p$Pb collisions~\cite{LHCb-PAPER-2018-048}. Therefore the $t_z
$ requirement produces data samples with a highly enriched prompt component and a negligible contribution from $b$ decays.  The resulting $\jpsi \pip \pim$ invariant mass spectra from $pp$, $p$Pb, and Pb$p$ collisions are shown in Fig.~\ref{fig:Fig1}. 

The $\jpsi \pip \pim$ mass distributions are fit to extract the ratio of $\theX$ to $\psitwos$ signal yields.  In the fit, the $\theX$ lineshape is represented by a Gaussian function, while the $\psitwos$ peak is represented by the sum of two Crystal Ball functions, with both low- and high-mass tails~\cite{CB}.  The background is studied by constructing the invariant mass spectrum of $\jpsi \pi^{\pm} \pi^{\pm}$ combinations using like-sign dipions, and is well represented by a third-order Chebychev polynomial in all data sets.  When fitting the $p$Pb and Pb$p$ samples, the $\theX$ and $\psitwos$ lineshapes including the $\theX$ mass are fixed to the values determined by fitting the relatively large $pp$ sample, while the $\psitwos$ mass, the signal yields, and the background parameters are allowed to float.  

The $\theX$ signal yields with their statistical uncertainties are determined to be $129\pm37$ and $71\pm39$ for the $p$Pb and Pb$p$ data sets, respectively.  The corresponding $\psitwos$ yields are $343\pm32$ and $191\pm30$ for the $p$Pb and Pb$p$ data sets.    Fit projections are shown overlaid on the data in Fig.~\ref{fig:Fig1}.  The statistical significance of the $\theX$ signal is estimated by calculating $\sqrt{-2\rm ln \frac{\mathcal{L}_{B}}{\mathcal{L}_{S+B}}}$ where $\mathcal{L}_{B}$ and $\mathcal{L}_{S+B}$ are the likelihoods under the background-only and signal-plus-background hypotheses~\cite{Wilks:1938dza}.  The resulting $\theX$ signal significance is $3.6\sigma$ for the $p$Pb data and $1.9\sigma$ for the Pb$p$ data. A systematic uncertainty on the fitting procedure is evaluated by changing the $\theX$ fit function to a relativistic Breit--Wigner convolved with a resolution function or a sum of two Crystal Ball functions, and allowing the $\theX$ mass to float.  The resulting variation in the ratio of $\theX$ to $\psitwos$ signal yields is taken as a systematic uncertainty, which is $5\%$ for the $p$Pb data and 27$\%$ for the Pb$p$ data.

Simulation is required to model the effects of the detector acceptance and the imposed selection requirements.  In the simulation, $\theX$ and $\psitwos$ particles are generated using \pythia~\cite{Sjostrand:2007gs} with a specific \lhcb configuration~\cite{LHCb-PROC-2010-056}, and embedded into the EPOS generator~\cite{Pierog:2013ria}, which simulates the environment produced in $p$Pb collisions. Decays of unstable particles are described by \evtgen~\cite{Lange:2001uf}.  The interaction of the generated particles with the detector, and its response, are implemented using the \geant toolkit~\cite{Allison:2006ve, *Agostinelli:2002hh} as described in Ref.~\cite{LHCb-PROC-2011-006}. The $\pt$ distributions of the simulated $\theX$ and $\psitwos$ decays are weighted to match distributions extracted from the data using the \sPlot method~\cite{splot} and the results of the fits in Fig.~\ref{fig:Fig1}.

The ratio of cross-sections $\sigma^{\theX}/\sigma^{\psitwos}$ times their branching fractions $\BR$ to $ \jpsi \pip \pim$ is given by

\begin{equation} 
\begin{aligned}
& \frac{\sigma^{\theX}}{\sigma^{\psitwos}} \times
\frac{\BR[\theX \rightarrow \jpsi \pip \pim]}{\BR[\psitwos \rightarrow \jpsi \pip \pim]} = \\
& \frac{N_{\theX} }{N_{\psitwos}} \times \frac{\epsilon^{\rm acc}_{\psitwos}}{\epsilon^{\rm acc}_{\theX}} \times \frac{\epsilon^{\rm trig}_{\psitwos}}{\epsilon^{\rm trig}_{\theX}} \times \frac{\epsilon^{\rm reco}_{\psitwos}}{\epsilon^{\rm reco}_{\theX}} \times  \left[\frac {\epsilon^{\mu^\pm \rm PID}_{\psitwos}}{\epsilon^{\mu^\pm \rm PID}_{\theX} } \right]^2  \times \left[\frac {\epsilon^{\pi^\pm \rm PID}_{\psitwos}}{\epsilon^{\pi^\pm \rm PID}_{\theX} }\right]^{2},
\end{aligned}
\end{equation} 

\noindent where $N_{\theX}/{N_{\psitwos}}$ is the ratio of signal yields returned by the fit, and the efficiency ratios are discussed below.

The ratio of LHCb's geometric acceptance for the daughter products $\epsilon^{\rm acc}_{\psitwos}/\epsilon^{\rm acc}_{\theX}$ is determined from simulation to be close to unity with a systematic uncertainty of 1$\%$, due to the uncertainty on the weights applied to the simulation to match the data.  The ratio of trigger efficiencies $\epsilon^{\rm trig}_{\psitwos}/\epsilon^{\rm trig}_{\theX}$ is determined from data to be consistent with unity within an uncertainty of $2\%$, using techniques described in Ref.~\cite{LHCb-PUB-2014-039}, where the uncertainty comes from statistical uncertainties on the data sample.  The ratio of reconstruction efficiencies $\epsilon^{\rm reco}_{\psitwos}/\epsilon^{\rm reco}_{\theX}$ is determined to be $0.67\pm0.12$ ($0.61\pm0.19$) for the $p$Pb (Pb$p$) data samples, where the uncertainty is due to the statistical uncertainty on the $\pt$ distributions of signals extracted from the data.  The deviation of this term from unity is due to the difference in the kinematics of $\theX$ and $\psitwos$ decays.  The dipions from $\psitwos\rightarrow\jpsi\pip\pim$ decays have masses between $\sim300$ and 600\mevcc, while the dipions from $\theX\rightarrow\jpsi\pip\pim$ decays are dominated by intermediate $\rho$ and $\omega$ states with higher mass and are reconstructed with a higher efficiency~\cite{LHCb-PAPER-2021-045}.  The ratios of muon and pion particle identification efficiencies, $\epsilon^{\mu^\pm \rm PID}_{\psitwos}/\epsilon^{\mu^\pm \rm PID}_{\theX}$ and $\epsilon^{\pi^\pm \rm PID}_{\psitwos}/\epsilon^{\pi^\pm \rm PID}_{\theX}$,  are determined using calibration samples of identified particles from the data to be consistent with unity, with uncertainties of 1$\%$ due to the finite size of those samples~\cite{LHCb-PUB-2016-021}.

\begin{figure}[t]
  \centering
 \includegraphics[width=0.9\linewidth]{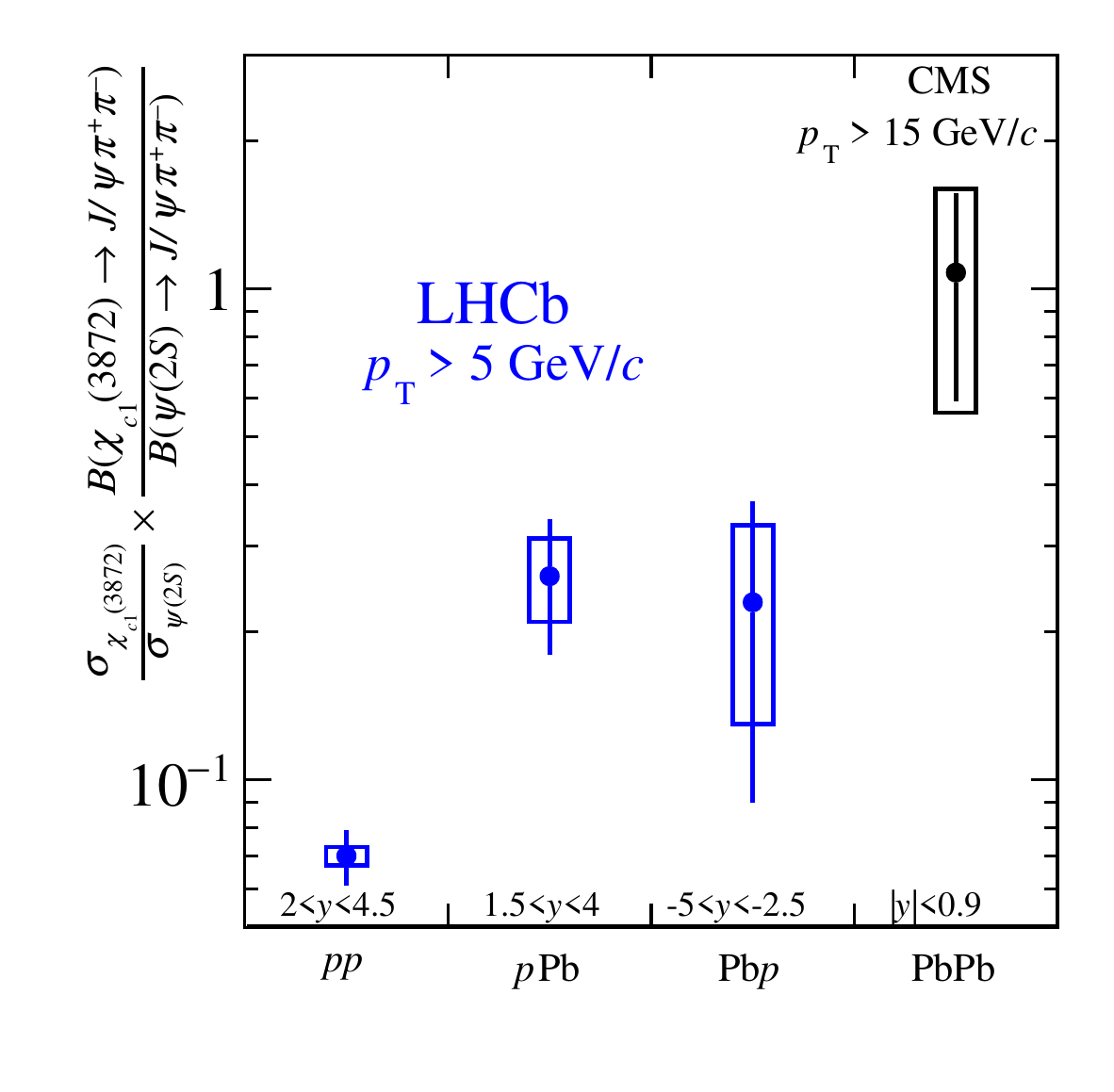}
  \caption{Ratio of $\theX$ to $\psitwos$ cross-sections in the $\jpsi \pip \pim$ decay channel, measured in $pp$~\cite{LHCb-PAPER-2020-023}, $p$Pb, Pb$p$, and PbPb~\cite{CMS:2021znk} data. The error bars (boxes) represent the statistical (systematic) uncertainties on the ratio. }
  \label{fig:Fig2}
\end{figure}

The resulting ratios of fiducial cross-sections times branching fractions are \mbox{$0.26\pm0.08\pm0.05$} and $0.23\pm0.14\pm0.10$ in the $p$Pb and Pb$p$ data samples, respectively, where the first and second uncertainties are statistical and systematic, respectively.  These ratios are shown in Fig.~\ref{fig:Fig2}, along with the ratio obtained from multiplicity-integrated LHCb data from $pp$ collisions~\cite{LHCb-PAPER-2020-023}, which has a value of $0.070\pm0.009\pm0.003$. CMS data from PbPb collisions is also included~\cite{CMS:2021znk}, which is measured over the rapidity interval $|y|<0.9$ and in a significantly higher $\pt$ range than in the LHCb measurements.  In this ratio, some effects that modify charm production in nuclear collisions, such as modification of the nuclear parton distribution function, largely cancel, leaving final-state effects as the dominant modification mechanism.  There is an increase in the ratio as the system size increases, which may be due to a combination of effects.  It has been observed that $\psitwos$ production is suppressed in $p$A collisions~\cite{PPG151, ALICEpPb, LHCb-PAPER-2015-058, PPG188,CMSpPbpsi2s,ALICEpPb2,Krista_psi2s,LHCb-PAPER-2023-024}, which would drive the ratio upwards even if no final-state effects modify $\theX$ production.  However, given that $pp$ collisions show a decreasing trend with multiplicity~\cite{LHCb-PAPER-2020-023}, the increase of the ratio may indicate that the hadronic densities achieved in the $p$Pb and Pb$p$ configurations allow quark coalescence to become the dominant mechanism affecting $\theX$ production.  

This ambiguity between $\psitwos$ suppression versus $\theX$ enhancement can be clarified by calculating the nuclear modification factor $R_{p \rm Pb}$. This factor is defined as the cross-section $\sigma_{p \rm A}$ measured in $p$Pb collisions divided by the cross-section $\sigma_{pp}$ measured in $pp$ collisions scaled by the number of nucleons in the nuclear beam, which is 208 for the Pb nuclei used in the LHC.  In this case, the ratios of cross-sections from $pp$ and $p$Pb collisions shown in Fig.~\ref{fig:Fig2}, along with the nuclear modification factor of $\psitwos$ (measured with relatively high precision in the dimuon channel in Ref.~\cite{LHCb-PAPER-2023-024}), can be used to find the $\theX$ nuclear modification factor via the equation


\begin{equation}
R_{p \rm A}^{\theX} = \frac{\sigma_{p \rm A}^{\theX}}{208\times\sigma_{pp}^{\theX}} = \frac{1}{208} \frac{\sigma_{p \rm A}^{\theX}}{\sigma_{pp}^{\theX}}\frac{\sigma_{p \rm A}^{\psitwos}}{\sigma_{pp}^{\psitwos}} \frac{\sigma_{pp}^{\psitwos}}{\sigma_{p \rm A}^{\psitwos}} = \newline R_{p \rm A}^{\psitwos} \frac{\sigma_{p \rm A}^{\theX}/\sigma_{p \rm A}^{\psitwos}}{\sigma_{pp}^{\theX}/\sigma_{pp}^{\psitwos}} .
\end{equation}

\begin{figure}[t]
  \centering
 \includegraphics[width=0.9\linewidth]{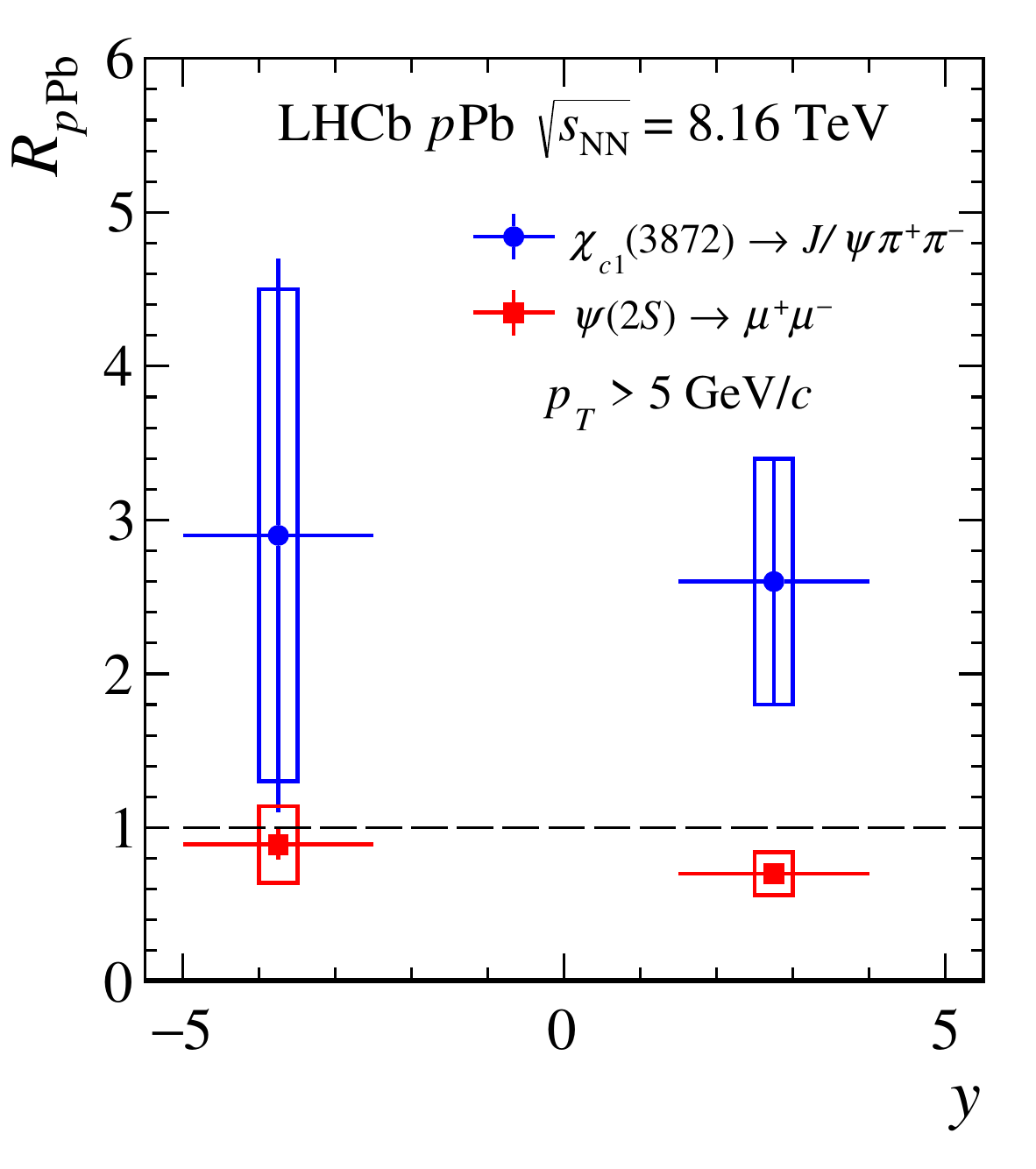}
  \caption{Nuclear modification factor $R_{p \rm Pb}$ for $\theX$ and $\psitwos$ hadrons~\cite{LHCb-PAPER-2023-024}.  The error bars (boxes) represent the statistical (systematic) uncertainties. }
  \label{fig:Fig3}
\end{figure}

The resulting nuclear modification factors are $2.6\pm0.8\pm0.8$ in $p$Pb and $2.9\pm1.8\pm1.6$ in Pb$p$, where the first and second uncertainties are statistical and systematic, respectively.  These results are shown as a function of rapidity in Fig.~\ref{fig:Fig3}, along with the $\psitwos$ measurement from Ref.~\cite{LHCb-PAPER-2023-024} for comparison.  An enhancement of $\theX$ production in $p$Pb collisions as compared to $pp$ collisions is seen, with significant uncertainties.  This could indicate that the enhancing effects of coalescence dominate $\theX$ production over the suppressing effects of breakup in $p$Pb collisions.  In the compact tetraquark interpretation of the $\theX$ structure, formation via coalescence could occur when a $c\bar{c}$ pair combines with two light quarks.  In $p$Pb collisions, the pseudorapidity density of produced charged particles is significantly higher than in $pp$ collisions~\cite{ALICE:2012xs,ATLAS:2015hkr}, providing an increased probability for quarks to overlap in position and velocity space and potentially coalesce.

In summary, the LHCb collaboration has produced the first measurements of $\theX$ production in $p$Pb collisions.  The increase of the ratio of cross-sections $\sigma^{\theX}/\sigma^{\psitwos}$ from $pp$ to $p$Pb to PbPb collisions may indicate that the exotic $\theX$ hadron experiences different dynamics in the nuclear medium than the conventional charmonium state $\psitwos$.  The nuclear modification factor $R_{p \rm A}$ shows that production of $\theX$ hadrons in $p$Pb collisions may be enhanced relative to $pp$ collisions, although significant uncertainties preclude drawing firm conclusions.  These first measurements of exotic hadron production in $p$Pb collisions can provide new constraints on the allowed configurations of quarks inside hadrons and on models of parton transport and hadronization in nuclear collisions.

\FloatBarrier

%% file: acknowledgements.tex
\section*{Acknowledgements}
%
%
\noindent We express our gratitude to our colleagues in the CERN
accelerator departments for the excellent performance of the LHC. We
thank the technical and administrative staff at the LHCb
institutes.
We acknowledge support from CERN and from the national agencies:
CAPES, CNPq, FAPERJ and FINEP (Brazil); 
MOST and NSFC (China); 
CNRS/IN2P3 (France); 
BMBF, DFG and MPG (Germany); 
INFN (Italy); 
NWO (Netherlands); 
MNiSW and NCN (Poland); 
MCID/IFA (Romania); 
MICINN (Spain); 
SNSF and SER (Switzerland); 
NASU (Ukraine); 
STFC (United Kingdom); 
DOE NP and NSF (USA).
We acknowledge the computing resources that are provided by CERN, IN2P3
(France), KIT and DESY (Germany), INFN (Italy), SURF (Netherlands),
PIC (Spain), GridPP (United Kingdom), 
CSCS (Switzerland), IFIN-HH (Romania), CBPF (Brazil),
and Polish WLCG (Poland).
We are indebted to the communities behind the multiple open-source
software packages on which we depend.
Individual groups or members have received support from
ARC and ARDC (Australia);
Key Research Program of Frontier Sciences of CAS, CAS PIFI, CAS CCEPP, 
Fundamental Research Funds for the Central Universities, 
and Sci. \& Tech. Program of Guangzhou (China);
Minciencias (Colombia);
EPLANET, Marie Sk\l{}odowska-Curie Actions, ERC and NextGenerationEU (European Union);
A*MIDEX, ANR, IPhU and Labex P2IO, and R\'{e}gion Auvergne-Rh\^{o}ne-Alpes (France);
AvH Foundation (Germany);
ICSC (Italy); 
GVA, XuntaGal, GENCAT, Inditex, InTalent and Prog.~Atracci\'on Talento, CM (Spain);
SRC (Sweden);
the Leverhulme Trust, the Royal Society
 and UKRI (United Kingdom).

%% file: Authorship_LHCb-PAPER-2023-026.tex
\centerline
{\large\bf LHCb collaboration}
\begin
{flushleft}
\small
R.~Aaij$^{35}$\lhcborcid{0000-0003-0533-1952},
A.S.W.~Abdelmotteleb$^{54}$\lhcborcid{0000-0001-7905-0542},
C.~Abellan~Beteta$^{48}$,
F.~Abudin{\'e}n$^{54}$\lhcborcid{0000-0002-6737-3528},
T.~Ackernley$^{58}$\lhcborcid{0000-0002-5951-3498},
B.~Adeva$^{44}$\lhcborcid{0000-0001-9756-3712},
M.~Adinolfi$^{52}$\lhcborcid{0000-0002-1326-1264},
P.~Adlarson$^{78}$\lhcborcid{0000-0001-6280-3851},
C.~Agapopoulou$^{46}$\lhcborcid{0000-0002-2368-0147},
C.A.~Aidala$^{79}$\lhcborcid{0000-0001-9540-4988}
% ...

{\footnotesize \it

% ...
$^{35}$Nikhef National Institute for Subatomic Physics, Amsterdam, Netherlands\\
% ...
$^{44}$Instituto Galego de F{\'\i}sica de Altas Enerx{\'\i}as (IGFAE), Universidade de Santiago de Compostela, Santiago de Compostela, Spain\\
% ...
$^{46}$European Organization for Nuclear Research (CERN), Geneva, Switzerland\\
% ...
$^{48}$Physik-Institut, Universit{\"a}t Z{\"u}rich, Z{\"u}rich, Switzerland\\
% ...
$^{52}$H.H. Wills Physics Laboratory, University of Bristol, Bristol, United Kingdom\\
% ...
$^{54}$Department of Physics, University of Warwick, Coventry, United Kingdom\\
% ...
$^{57}$School of Physics and Astronomy, University of Glasgow, Glasgow, United Kingdom\\
$^{58}$Oliver Lodge Laboratory, University of Liverpool, Liverpool, United Kingdom\\
% ...
$^{66}$Syracuse University, Syracuse, NY, United States\\
% ...
$^{78}$Department of Physics and Astronomy, Uppsala University, Uppsala, Sweden, associated to $^{57}$\\
$^{79}$University of Michigan, Ann Arbor, MI, United States, associated to $^{66}$\\
% ...
}
\end{flushleft}